\documentclass[12pt,preprint]{aastex}

\begin{document}


\title{A robust correlation between growth rate and amplitude of solar
  cycles: consequences for prediction methods}

\author{R. Cameron}\email{cameron@mps.mpg.de}

\and

\author{M. Sch\"ussler}\email{schuessler@mps.mpg.de}

\affil{Max-Planck-Institut f\"ur Sonnensystemforschung, 
   37191 Katlenburg-Lindau, Germany}

\begin{abstract}
We consider the statistical relationship between the growth rate of
activity in the early phase of a solar cycle with its subsequent
amplitude on the basis of four datasets of global activity indices (Wolf
sunspot number, group sunspot number, sunspot area, and 10.7-cm radio
flux). In all cases, a significant correlation is found: stronger cycles
tend to rise faster. Owing to the overlapping of sunspot cycles, this
correlation leads to an amplitude-dependent shift of the solar minimum
epoch. We show that this effect explains the correlations underlying
various so-called precursor methods for the prediction of solar cycle
amplitudes and also affects the prediction tool of Dikpati et al. (2006)
based upon a dynamo model.  Inferences as to the nature of the solar
dynamo mechanism resulting from predictive schemes which (directly or
indirectly) use the timing of solar minima should therefore be treated
with caution.

\end{abstract}


\section{Introduction} 
Solar activity is the driver of space weather, which has practical
consequences for human activities in space. This is one reason why the
search for methods to predict its (short-term and long-term) future
levels has found much interest in the literature. Further motivation
arises from the potential implications for understanding the {\em
origin} of solar activity: a reliable method of predicting the amplitude
of future solar cycles could provide a constraint on dynamo models. On
the other hand, the converse proposition is not necessarily true, as has
been pointed out by \citet{Bushby:Tobias:2007}: the nonlinear dynamics
of the dynamo might be such as to make mid- to long-term prediction
impossible even if an almost perfect physical understanding of the
dynamo mechanism is achieved.

In many cases, recipes for prediction are inferred from correlations
found in historical records of measured quantities, which are (directly
or indirectly) related to solar activity. As illustrated by
\citet[][cf. his Fig. 14.2]{Wilson:1994} and \citet[][cf. their
Fig.~6]{Lantos:Richard:1998}, the success of most methods in actually
{\em predicting} the unknown amplitude of a future cycle is rather
disappointing. Nevertheless, the correlations between several
`precursors', i.e., quantities measured during the descending or minimum
phase of a cycle and the amplitude of the subsequent cycle
\citep[e.g.,][]{Hathaway:etal:1999, Schatten:2003} might have a
non-random origin and thus call for a physical explanation. This could
have implications for dynamo models.

In this paper, we consider the effect of the overlapping of solar cycles
in combination with their asymmetric shape on the correlations between
precursors and following cycle amplitudes. In this connection, the
important aspect of the asymmetry is the difference of the
amplitude-dependent ascent rate in the early cycle phase (related to the
so-called Waldmeier effect) compared to the decay rate near the end of a
cycle. Since sunspot cycles overlap for typically 2 to 3 years
\citep{Harvey:1992a}, this asymmetry affects the timing of the activity
minima, which are pivotal epochs for most precursor methods. We show
that these effects can explain the correlations upon which such methods
are based, without necessarily implying a direct physical connection
between the precursor quantity and the following cycle. We also show
that the essence of the Waldmeier effect, i.e., that stronger cycles
tend to show a faster rise of activity levels during their ascending
phase than weaker cycles, is a robust property present in all activity
indices.  On this basis, we explain how cycle asymmetry and cycle
overlapping may also affect the dynamo-based prediction method of
\citet{Dikpati:Gilman:2006}, in spite of recent claims to the contrary
\citep{Dikpati:etal:2008}.

\section{Overlapping asymmetric cycles and precursors}

Precursor methods are based upon the existence of a correlation between
some physical quantity measured during the descending or minimum phase
of a cycle and the amplitude of the following cycle. The correlation is
established by considering historical records of data, the longest of
which is the record of sunspot numbers. Assuming a non-random origin of
such a correlation, there are three possible explanations for
its existence:
\begin{enumerate}
\item The precursor is a feature of the {\em old} cycle that represents
  or is related to an input quantity for the dynamo process, so that its
  magnitude directly affects the amplitude of the next cycle. For
  example, it has been suggested that the strength of the polar field
  during solar minimum is a measure of the poloidal field from which the
  toroidal field for the next cycle is generated
  \citep[e.g.,][]{Schatten:etal:1978, Choudhuri:etal:2007}.
\item The precursor quantity represents an early manifestation of the
  {\em new} cycle, which already affects the high latitudes of the Sun but
  does not yet produce sunspots. The notion of such an `extended solar
  cycle' \citep{Wilson:etal:1988} is supported by observations of
  coronal activity \citep{Altrock:1997}, ephemeral magnetic regions
  \citep{Harvey:1994a}, and zonal flows \citep{Howe:etal:2006,
  Altrock:etal:2008}.
\item If the definition of the precursor quantity directly or indirectly
  depends on the timing of the solar minimum, the amplitude-dependent
  shift of the minimum due to the overlap of asymmetric cycles can lead
  to a correlation of the precursor with the amplitude of the new cycle
  \citep[][see their Fig.~10]{Cameron:Schuessler:2007}. This does not
  necessarily involve a physical connection between the precursor and
  the dynamo process, in principle even permitting the prediction of
  random cycle amplitudes.
  
\end{enumerate}
The third possibility arises from the observation that information about
the height of the maximum of a cycle is already contained in its early
rise phase \citep{Waldmeier:1935,Waldmeier:1955}. In fact, the shapes of
most historical solar cycles can reasonably well be described by simple
functions containing a few parameters
\citep[e.g.,][]{Hathaway:etal:1994,Li:1999}, so that an estimate of the
further development of a cycle is possible a few years after sunspot
minimum \citep{Waldmeier:1936, Elling:Schwentek:1992}. Since cycles
overlap, i.e., the rise of the new cycle starts already when the decay
of the old cycle is still ongoing, the timing of the cycle minimum (the
epoch when the sum of the activities of both cycles is minimal) and its
height are affected by the amplitude of the new cycle: the minimum
occurs earlier and is higher for a subsequent strong cycle than for a
weak cycle. In fact, stronger cycles tend to be preceded by shorter
cycles \citep[earlier minima,][]{Solanki:etal:2002b} with higher minimum
activity levels. Both quantities can be considered as precursors with
correlation coefficients with the amplitude of the next cycle of about
0.7 \citep{Hathaway:etal:1999} .

\citet{Cameron:Schuessler:2007} have noted that the amplitude-dependent
shift of the minimum epoch is enhanced by the amplitude-dependent
{\em asymmetry} of solar cycles: strong cycles tend to grow faster in activity
than weak cycles, while the decay rate in the late descending phase is
largely independent of the cycle amplitude.  This is related to, but not
identical to, the so-called `Waldmeier effect' \citep{Waldmeier:1935}
which is often stated in the form: the time between cycle minimum and
maximum is shorter for stronger cycles
\cite[e.g.,][]{Hathaway:etal:2002}. However, the crucial quantity for
the minimum shift is not the time between minimum and maximum but the
{\em steepness} of the activity rise (decay) in the initial (late) phase
of the cycle, which can be determined independently of the timing of the
activity minima and maxima by considering the slope of the activity
curve. We show in the next section that the correlation between initial
growth rate and cycle amplitude is a very robust feature shown by all
available global activity indices.

\section{The robust amplitude-dependent cycle asymmetry}

\citet{Hathaway:etal:2002} found that the `classical' Waldmeier effect,
i.e., the correlation between the rise time (time interval between
minimum and maximum) and the cycle amplitude, is by a factor of about
two weaker for the group sunspot numbers \citep{Hoyt:Schatten:1998} than
for the Wolf sunspot numbers. For both datasets, the historically
strongest cycles (numbers 18, 19, 21, and 22) do not fit the linear
relationship very well, showing somewhat long rise times. This weakening
of the correlation can be understood by the shift in the timing of the
minima: stronger, rapidly growing cycles lead to early preceding minima,
thus increasing the time between minimum and maximum. This effect
becomes even more pronounced in the analysis of sunspot area data by
\citet{Dikpati:etal:2008} owing to the fact that the sunspot areas for
cycles 21, 22 and 23 reach a slightly higher second `Gnevyshev maximum'
\citep{Gnevyshev:1967} a few years after the first maximum coinciding
with the sunspot number maximum. It is not surprising that determining
the rise time in terms of these later maxima destroys the correlation
with the cycle amplitude.

As we have pointed out in the previous section, the relevant quantity
for the minimum shift of overlapping cycles is not the time interval
between minimum and maximum but the {\em steepness} of the activity
growth (decay) in the initial (late) phase of the cycle.  An inspection
of Fig.~5 of \citet{Hathaway:etal:2002} and of Fig.~1 of
\citet{Dikpati:etal:2008} already suggests that the different datasets
do not differ strongly in this respect.  In order to give a quantitative
account, we need to measure the steepness without reference to the
epochs of maximum and minimum; in particular, the latter is already
affected by the correlations that we wish to study. For instance,
\citet{Lantos:2000} considered the slope of the sunspot number curve at
the inflexion point during the ascending part of cycles 9 to 22 and
found a correlation coefficient of $r=0.88$ with the cycle amplitude,
while the (anti)correlation with the rise time from minimum to maximum
only yields $\vert r\vert=0.61$. In our case, we need to consider
specifically the early rise and late decay phases, so that we estimate
the rise and decay rates by determining the time required for the
activity index under consideration to cover a fixed interval of
values. This interval is chosen such that it is not significantly
affected by the cycle overlap. The rise and decay rates are then defined
as the ratios of the value intervals and the corresponding rise and
decay times, respectively.

\begin{figure}
\epsscale{1.0}
\plottwo{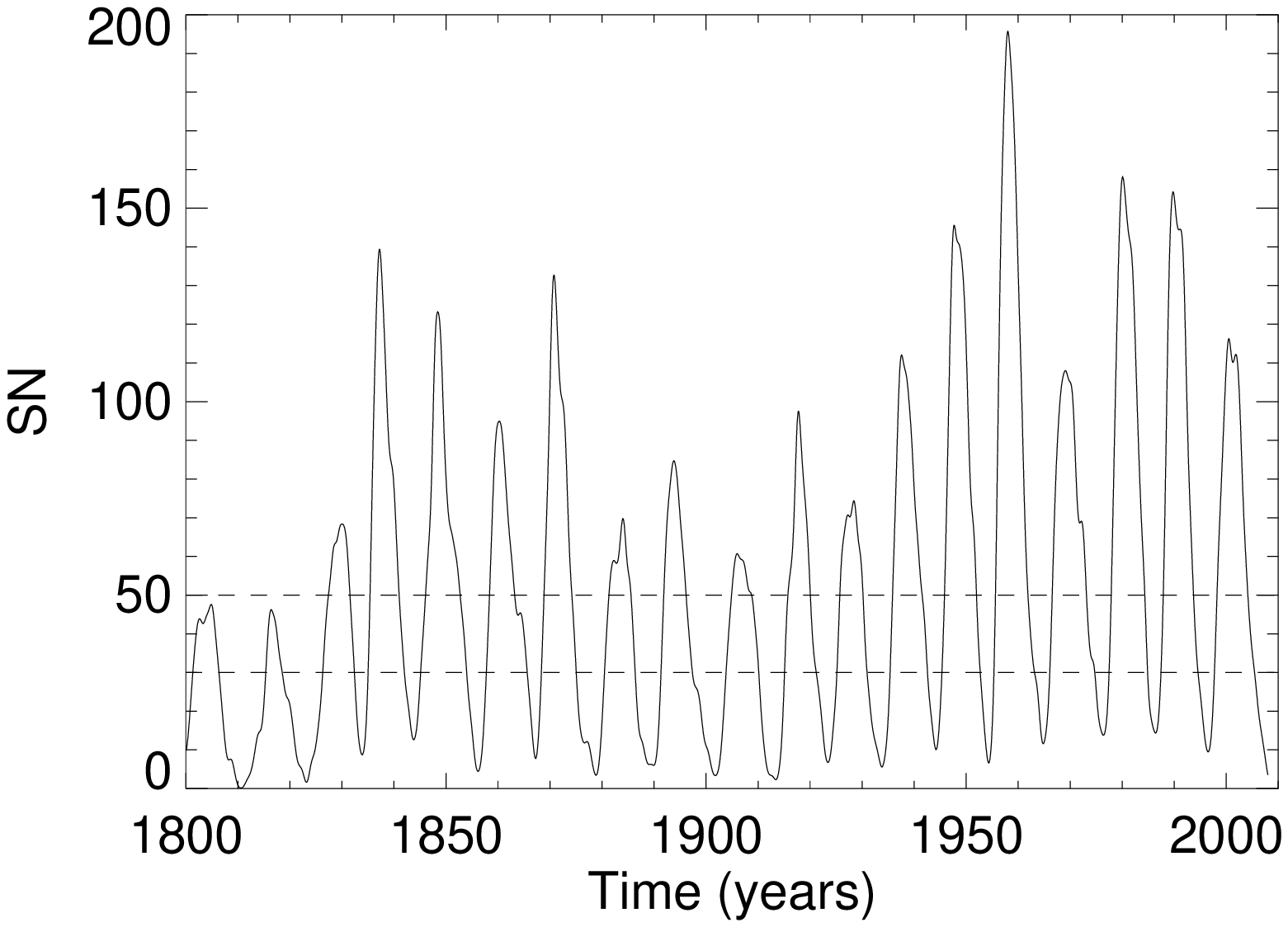}{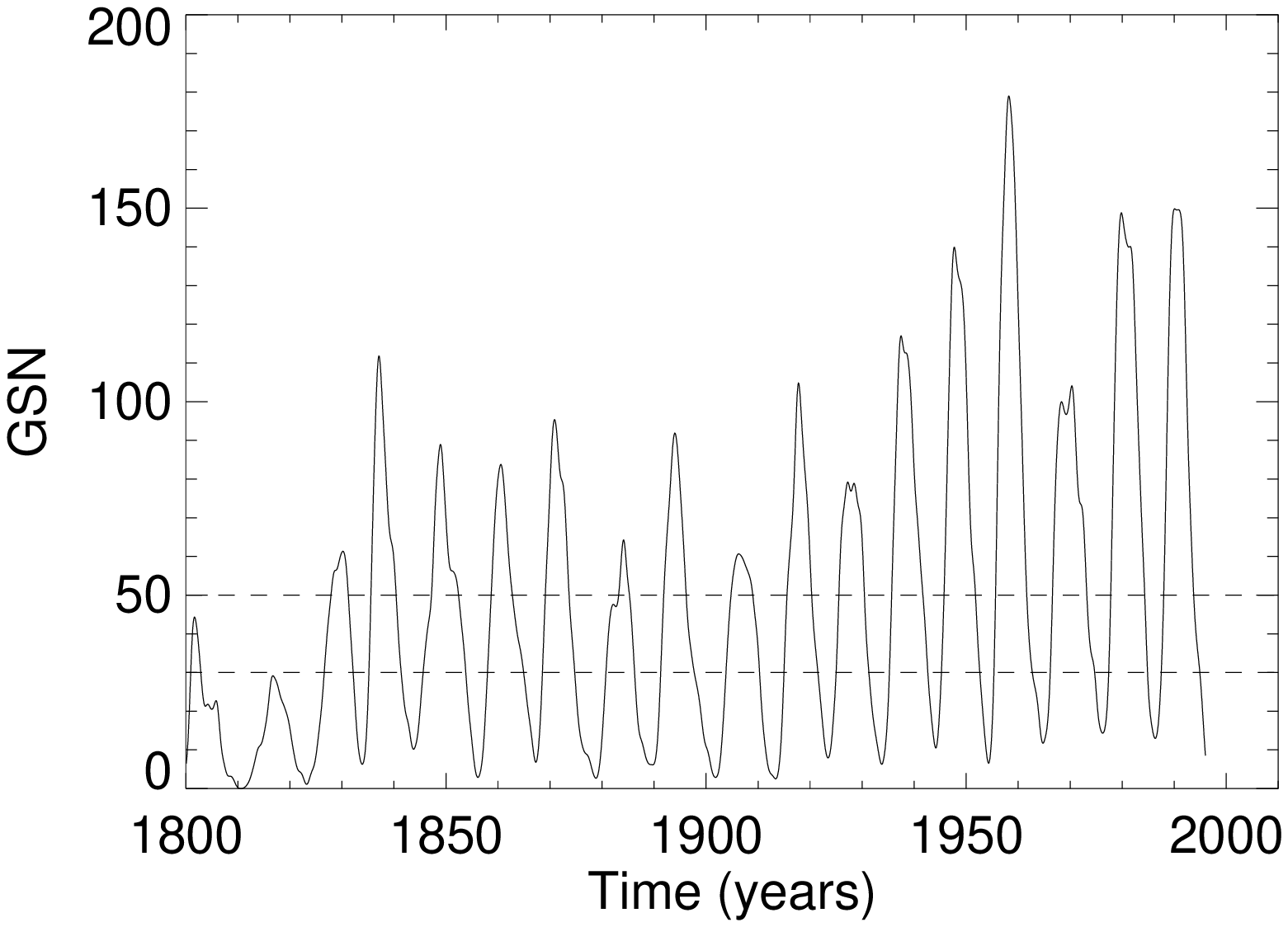}
\plottwo{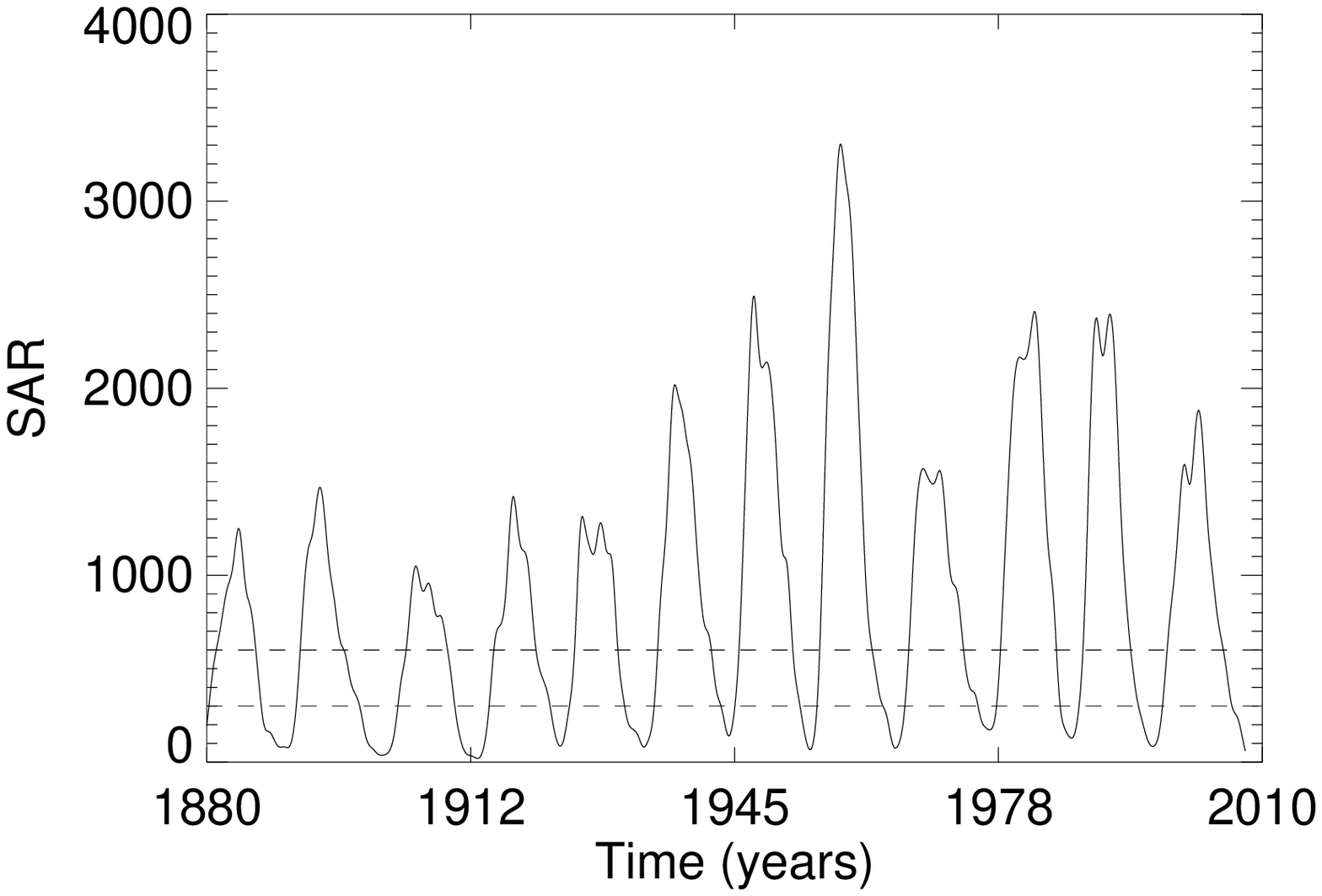}{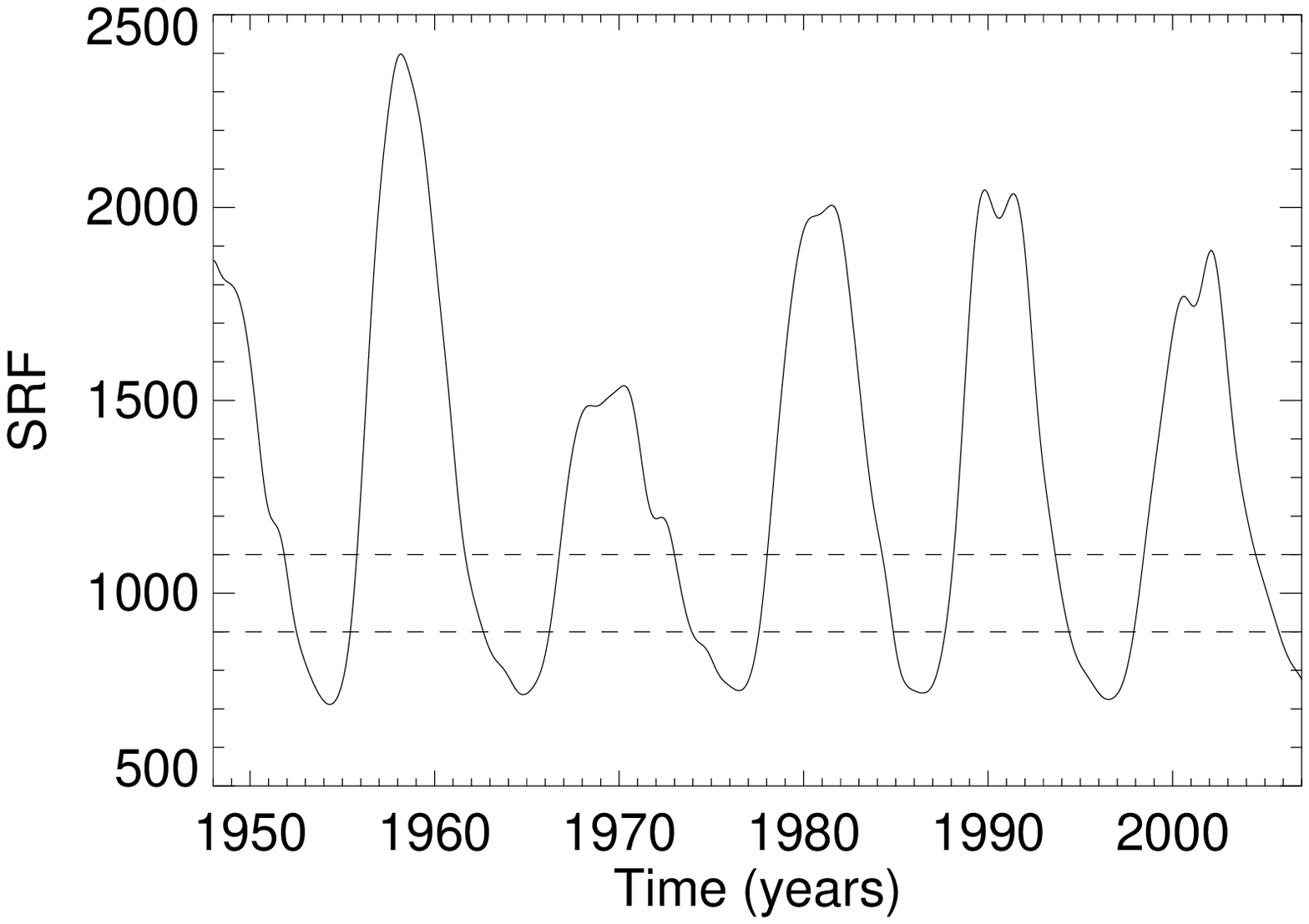}
\caption{Data sets used in this study: Wolf sunspot number (SN, top
 left), group sunspot number (GSN, top right), sunspot area (SAR, bottom
 left), and 10.7-cm solar radio flux (SRF, bottom right). SN and GSN are
 considered from cycle no. 7 onward (after the Dalton minimum). The
 dashed horizontal lines indicate the intervals used for the
 determination of the rise and decay rates.}
\label{fig_data}
\end{figure}

We have analyzed four datasets: monthly Wolf sunspot
numbers\footnote{http://sidc.oma.be} (SN) for cycles 7 to 23, monthly
group sunspot
numbers\footnote{http://www.ngdc.noaa.gov/stp/SOLAR/ftpsunspotnumber.htm}
\citep[GSN,][]{Hoyt:Schatten:1998} for cycles 7 to 22, sunspot areas
\citep[SAR,][]{Balmaceda:etal:2005} for cycles 12 to 23, and 10.7-cm
solar radio
flux\footnote{http://www.drao.nrc.ca/icarus/www/sol\_home.shtml} (SRF)
for cycles 19 to 23. While we have used the full available data sets for
SRF and SAR, the SN and GSN data are only considered from cycle 7 on
because a) we wish to compare the results for the different datasets
during roughly the same period of time, and b) for very low-amplitude
cycles like those during the Dalton minimum and the first cycles after
the Maunder minimum we cannot use the same intervals for the
determination of the rise and decay times.  Fig.~\ref{fig_data} shows
the datasets, also indicating the intervals chosen for the determination
of growth and decay rates: 30--50 (SN, GSN), 300--600~microhemispheres
(SAR), and 900--1100~sfu (SRF). All data have been smoothed by a Gaussian
with a FWHM of 1 year.

\clearpage

\begin{figure}
\epsscale{.9}
\plottwo{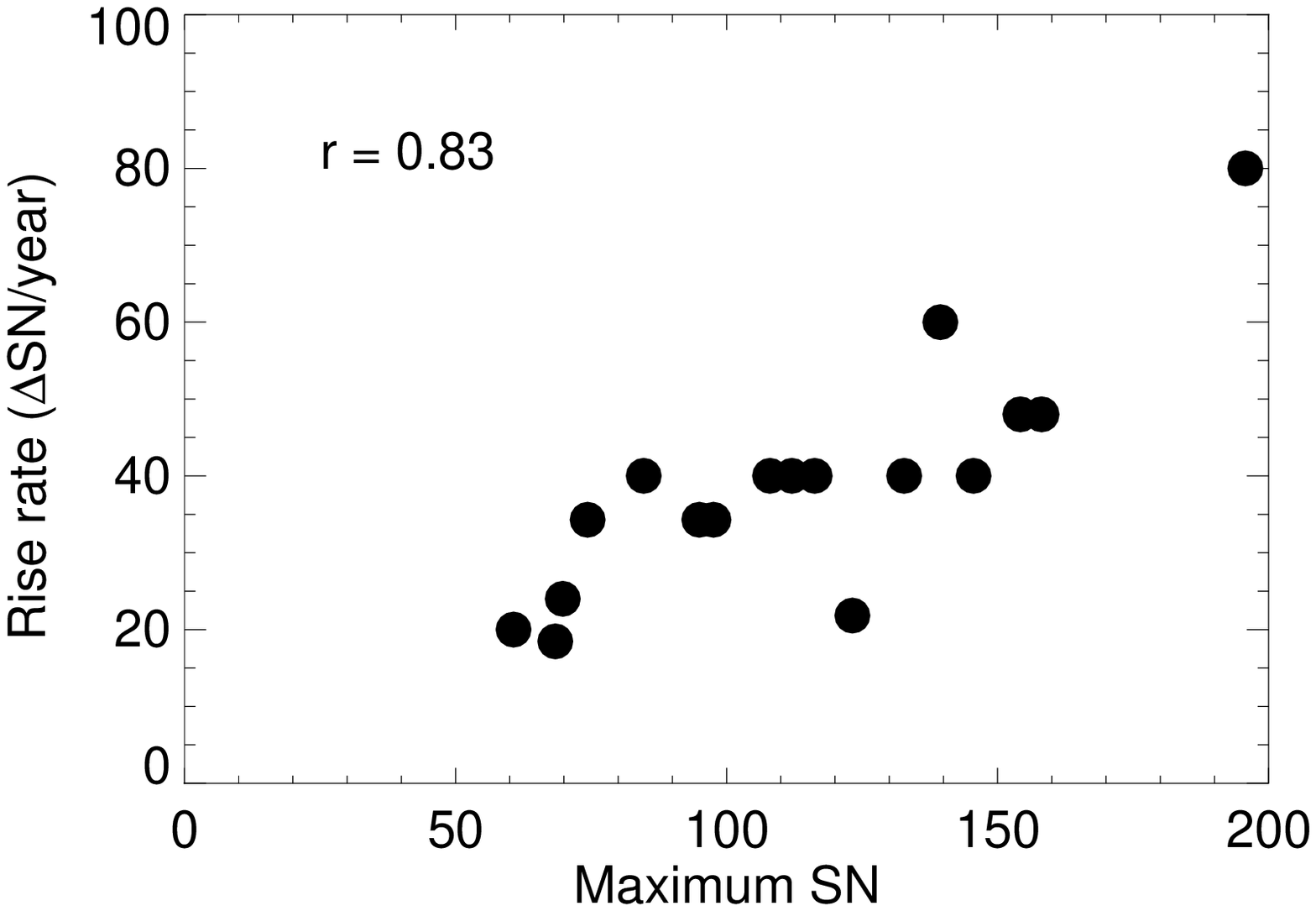}{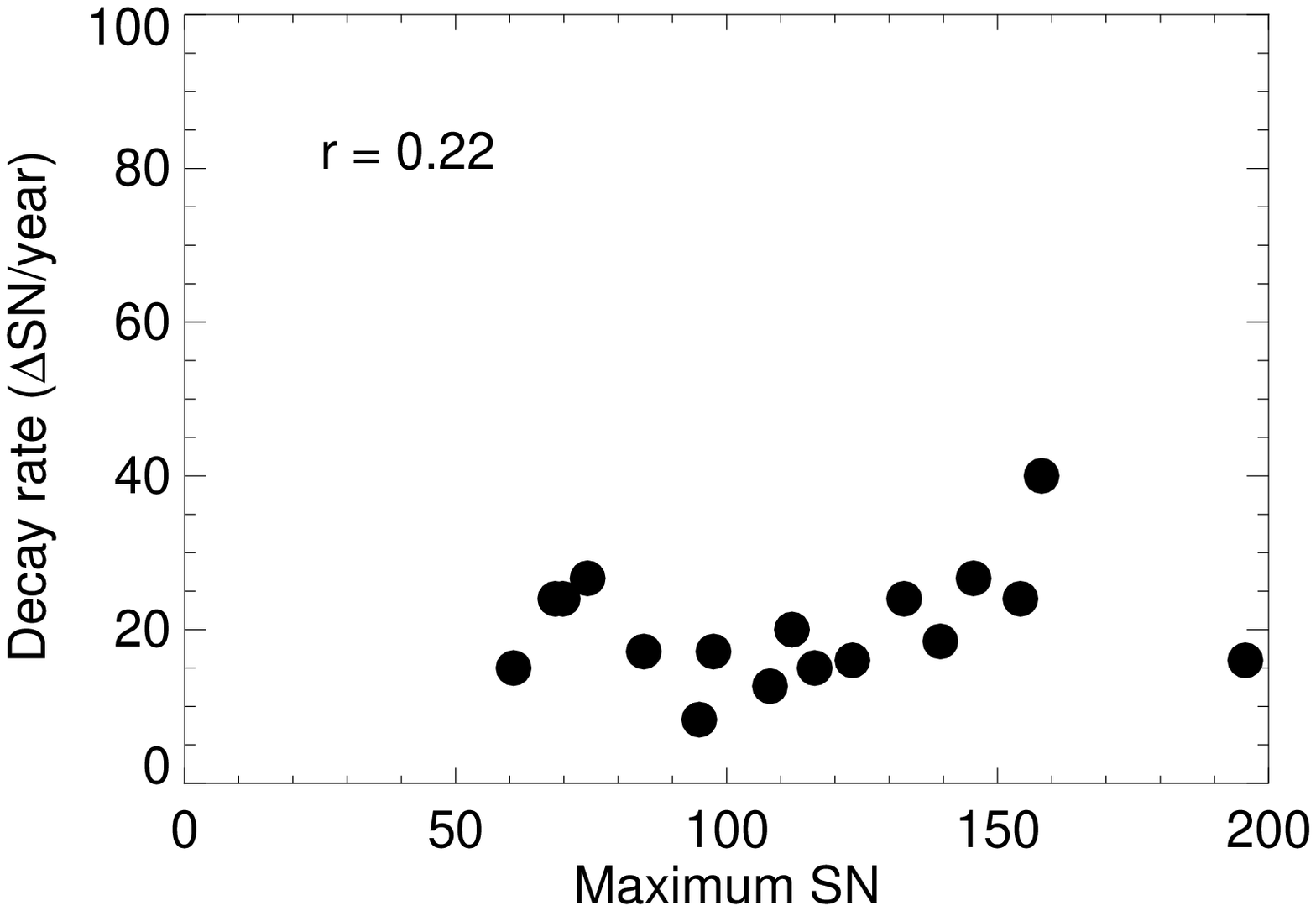}
\plottwo{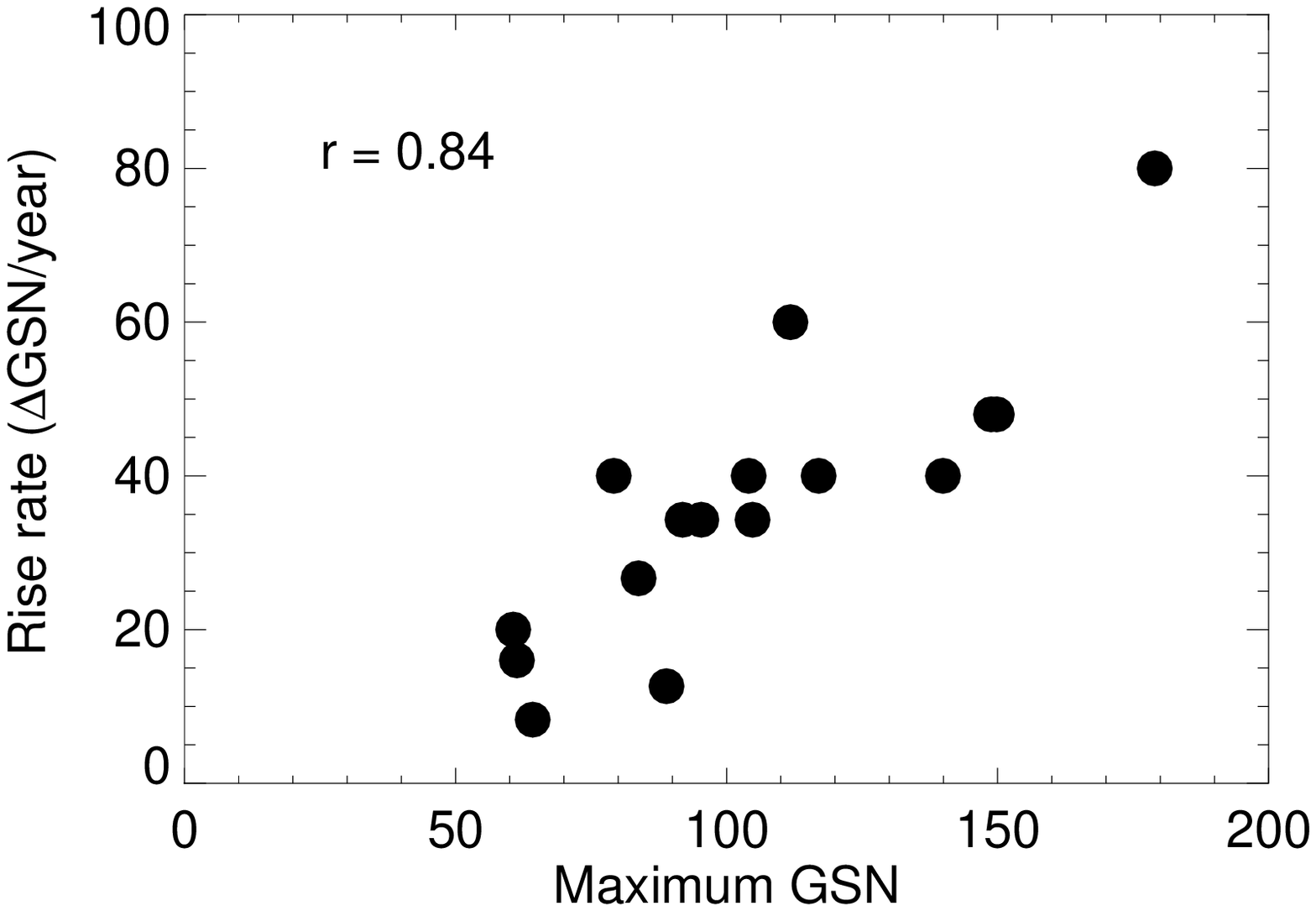}{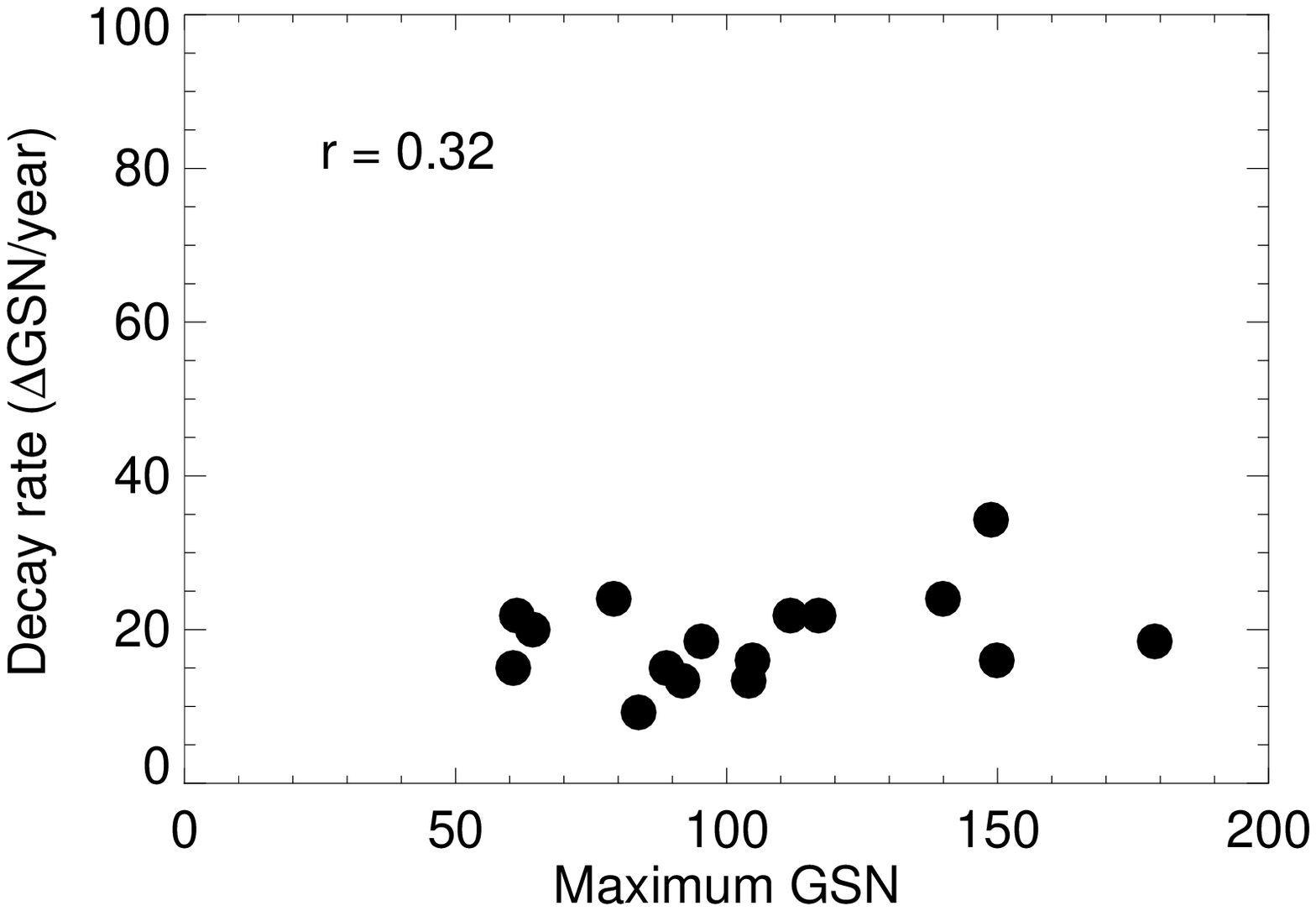}
\plottwo{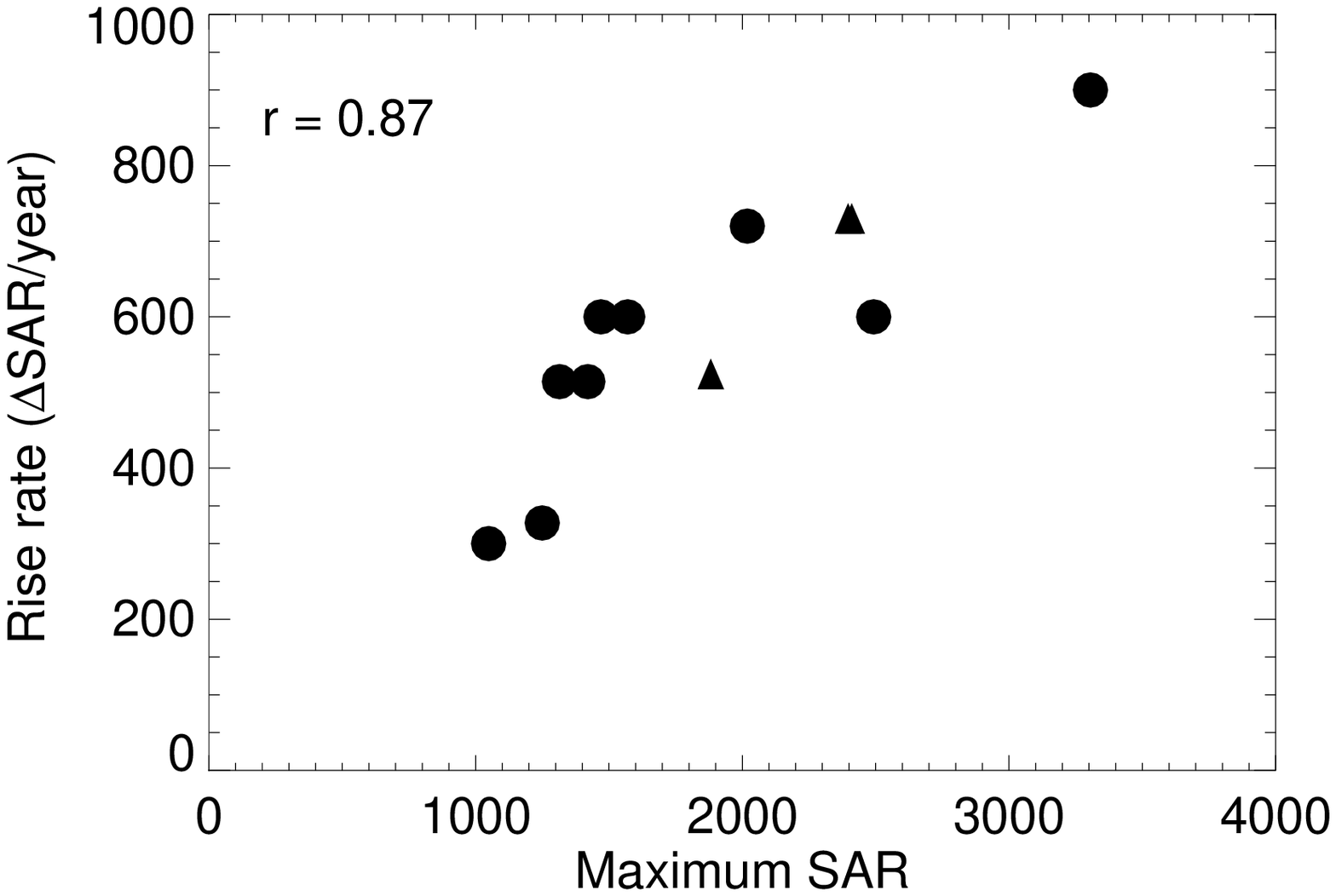}{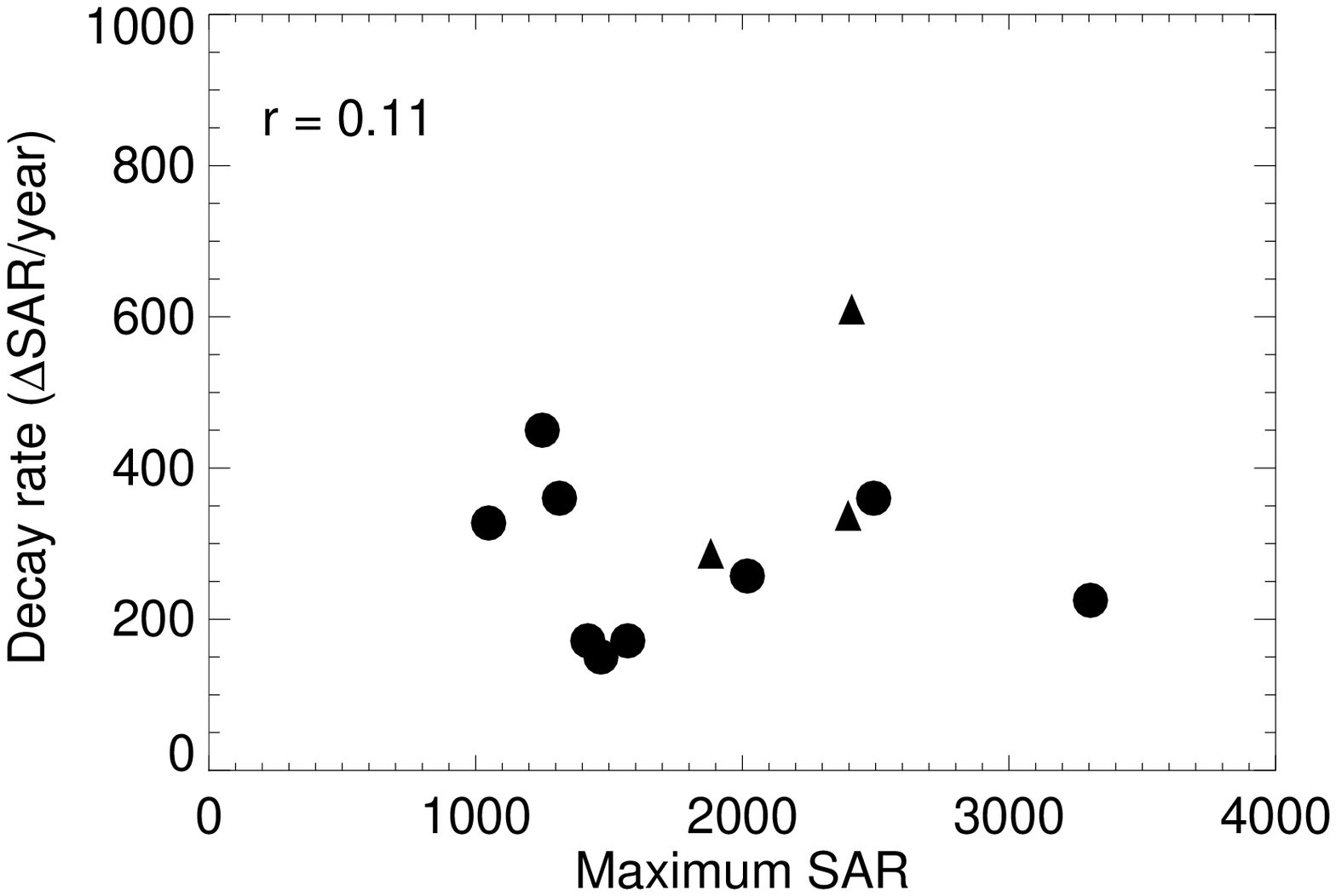}
\plottwo{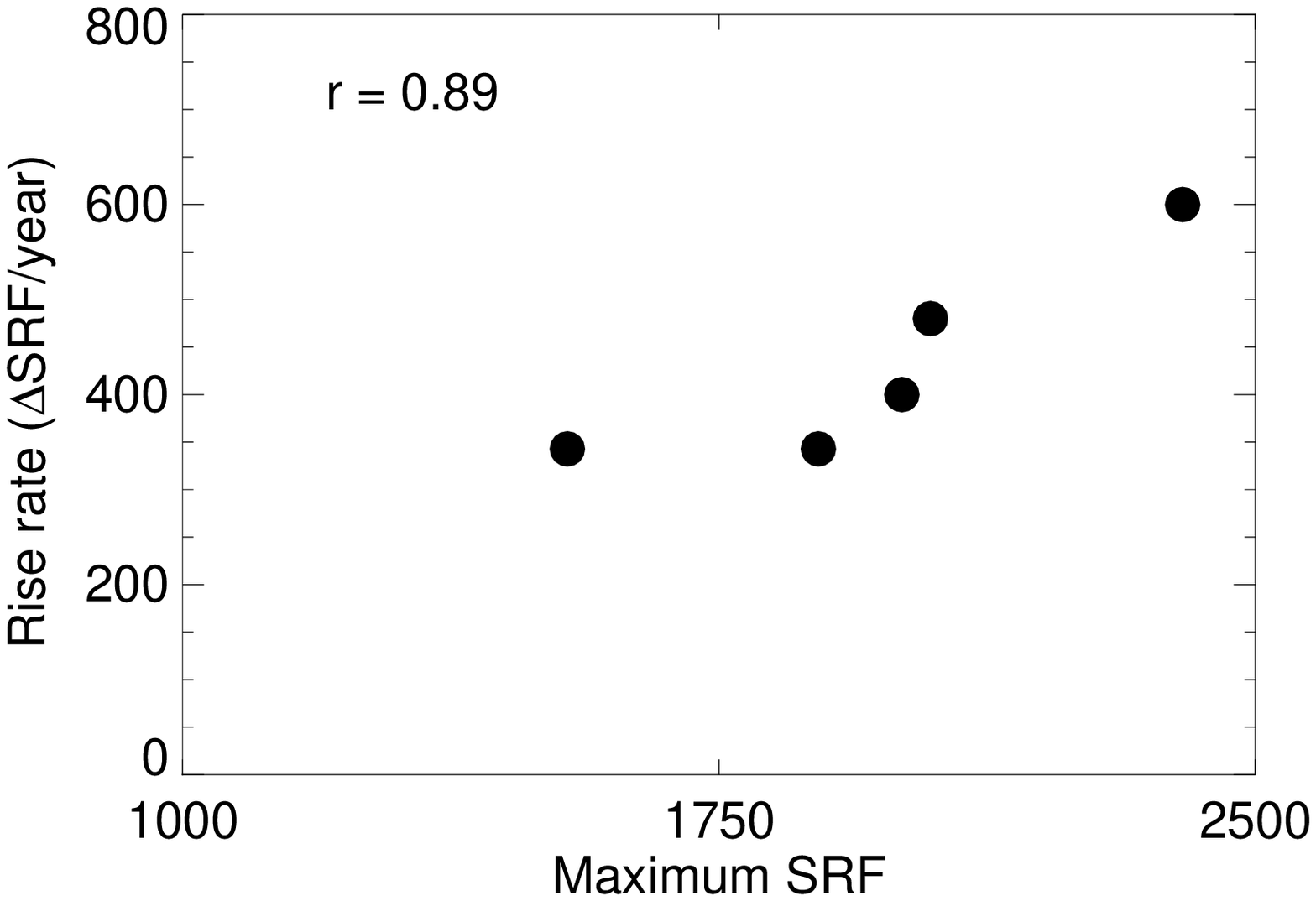}{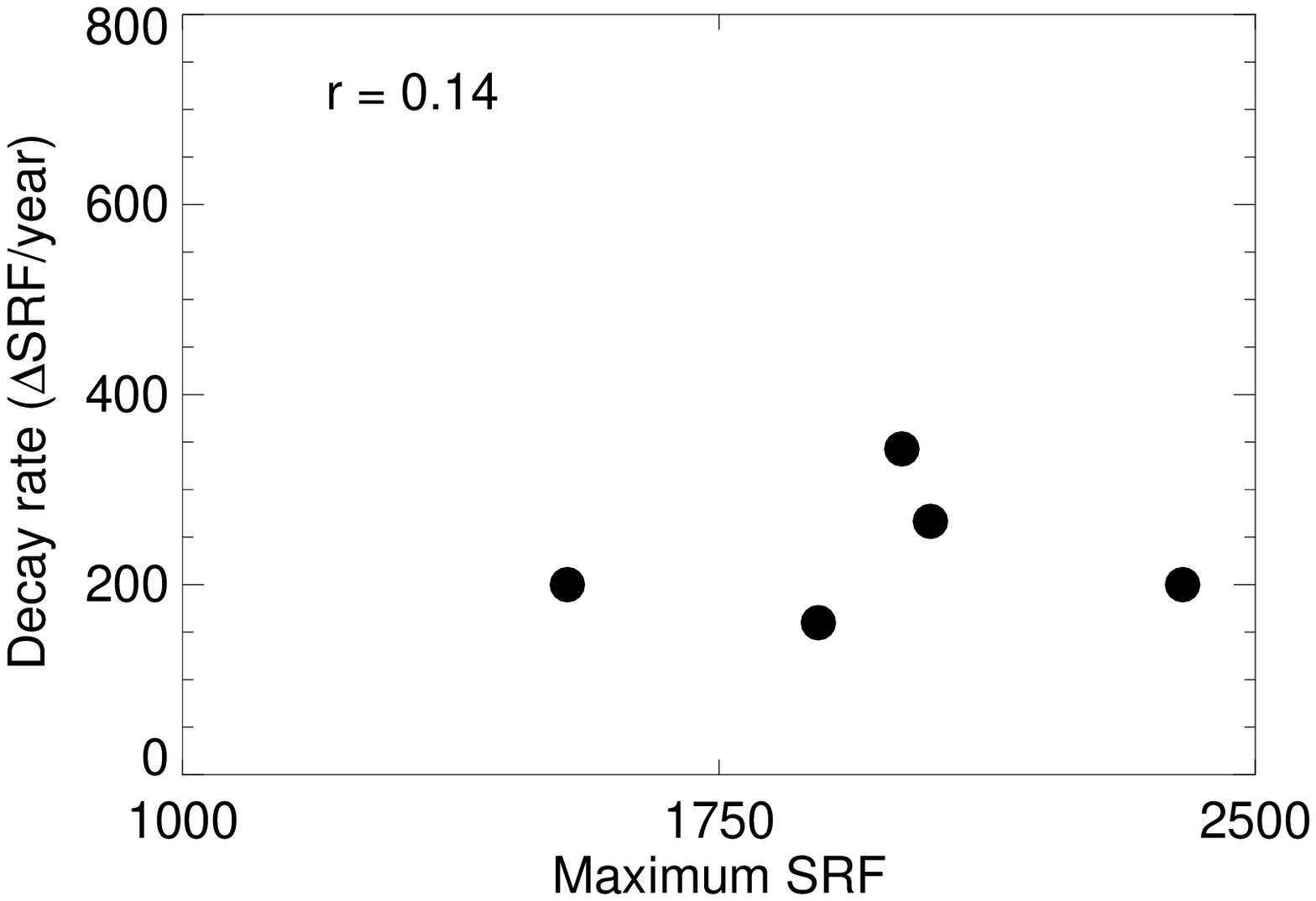}
\caption{Scatter diagrams of rise rates (left panels) and decay rates
(right panels) versus cycle amplitude for four sets of solar activity
indices: Wolf sunspot number (SN, top row), group sunspot number (GSN,
second row), sunspot area (SAR, third row, cycles 21--23 indicated by
triangles, two of which nearly coincide on the left-hand panel), and
10.7-cm solar radio flux (bottom row). The corresponding linear
correlation coefficients are given. }
\label{fig_corr}
\end{figure}
\clearpage

Scatter diagrams of the resulting rise and decay rates with the cycle
amplitudes are shown in Fig.~\ref{fig_corr}, together with the
corresponding correlation coefficients. The figure demonstrates that
high correlation coefficients ($r=0.83\dots0.89$) between rise rate and
cycle amplitude exist for all activity indices. Assuming a Gaussian
distribution, the corresponding significance levels are between 95\%
(SRF) and over 99\% (SN, GSN, SAR). In particular, the group sunspot
numbers and the sunspot areas exhibit an even higher correlation than
the Wolf sunspot numbers. Note that cycles 21--23 (indicated by
triangles in the third row of Fig.~\ref{fig_corr}), which, owing to
their higher {\bf Gnevyshev} maxima in the SAR data, largely destroyed the
correlation of the `classical' Waldmeier effect in the analysis of 
\citet{Dikpati:etal:2008}, nicely fit the correlation for the
growth rate.  On the other hand, the decay rates show no
significant correlation with the cycle amplitude, so that we find an
{\em amplitude-dependent asymmetry} between the rise rates early in a cycle
and the decay rates in the late phase. This asymmetry acts to make
the behavior near sunspot minimum rather insensitive to the amplitude of
the old cycle but sensitive to that of the new.

We have checked how strongly the correlation coefficients depend on the
chosen intervals for the determination of the growth and decay rates, and
found them to be rather insensitive. For instance, in the case of the growth
(decay) rates for the SN data, we have $r=0.79\,(0.05)$ for the interval
$[25,45]$, $r=0.82\,(0.22)$ for $[30,50]$ (the interval chosen for
Fig.~\ref{fig_corr}), $r=0.85\,(0.42)$ for $[35,55]$, and
$r=0.88\,(0.54)$ for $[40,60]$. Similar results are found for the other
datasets.


\section{Implications for precursor methods}

The overlapping of cycles with an amplitude-correlated asymmetry leads
to a dependence of the minimum epoch on the amplitude of the following
cycle: since the initial activity of a large-amplitude cycle increases
faster while the decline in the decay phase is not or less
amplitude-dependent, the minimum level of activity between overlapping
cycles occurs the earlier the higher the amplitude of the following
cycle. Such a minimum shift affects the value of any precursor quantity
which is defined (directly or indirectly) with regard to the minimum
epoch; in the extreme, it may completely explain the predictive power of
the precursor. \citet{Cameron:Schuessler:2007} have demonstrated this
for the case of a very simply defined precursor, namely, the activity
level (such as sunspot number or area), three years before minimum. The
earlier the minimum occurs (i.e., the stronger the new cycles is), the
higher is the precursor value, simply because it is measured at an
earlier time when the activity of the old cycle had not declined as
much. This explains why the method can, in principle, even predict
cycles with random amplitudes. It is important to note, however, that
such methods always require that the minimum epoch {\em is already
known}. For a real prediction (as opposed to a postdiction of past
cycles) such a method can be applied only some time into the new cycle,
when proper averaging can be performed to define the minimum time in
view of the strong fluctuations of activity around minimum
\citep[cf.][]{Harvey:White:1999b}.

Precursors that are defined (directly or indirectly) in relation to the
timing of the minimum are affected by the amplitude-dependent minimum
shift, which thus may (partly or fully) explain their correlation with
the amplitude of the next cycle (i.e., their predictive power). Among
the various precursor methods affected we mention just a few examples:
geomagnetic activity before and around solar minimum
\citep[e.g.,][]{Ohl:1966, Legrand:Simon:1981, Lantos:Richard:1998,
Kane:2007}, sunspot activity around minimum \citep{Hathaway:etal:1999},
length of the preceding sunspot cycle \citep{Kane:2008}, skewness of the
preceding cycle \citep{Ramaswamy:1977,Lantos:2006}, as well as
properties of the decay phase of the preceding cycle
\citep{Podladchikova:etal:2008}.

\section{Implications for the model of Dikpati et al.}

\citet{Cameron:Schuessler:2007} suggested that at least part of the
predictive skill of the model of \citet{Dikpati:etal:2006} and
\citet{Dikpati:Gilman:2006}, henceforth referred to as the DGT model,
may result from the overlap of asymmetric cycles in the sunspot area
data from which their source term is derived. We have shown above that
the underlying effect, i.e., the amplitude-dependent asymmetry between
the growth and decay phases, is present in all global activity datasets,
including the sunspot area data used in the DGT model.

The effect of this crosstalk between cycles on the DGT model results
from the assumption of a fixed speed and range of latitude drift of the
source term in the course of a cycle. The shortening of the cycle length
(defined as the time between minima) by the shift of the minimum
preceding a strong cycle then leads to larger source amplitudes (sunspot
areas) assigned to the low-latitude part of the artificial butterfly
diagram used in the DGT model. As a consequence, the magnetic flux
crossing the equator as the potentially relevant quantity for the dynamo
amplitude of the next cycle is higher, while the opposite is true for a
weak following cycle. In this way, the correlation in the data survives
the artificial stretching/compressing of cycles carried out in the DGT
model in order to obtain a constant cycle period. In fact,
\citet{Dikpati:etal:2008b} found that, in their model, the flux crossing
the equator during a cycle is significantly correlated ($r=0.76$) with
the amplitude of the next cycle. This is in accordance with Fig. 8a in
\citet{Dikpati:Gilman:2006}, which demonstrates that the memory in their
model does not extend much longer than one cycle, but it is in contrast
to the claim of \citet{Dikpati:etal:2008b} that it takes 17-21 years for
the poloidal surface field to reach the tachocline and generate the new
toroidal field, so that 2-3 preceding cycles contribute to the toroidal
flux of a cycle. {\bf A stronger diffusive coupling between surface and
tachocline than indicated by the explicit value of their magnetic
diffusivity (for example, due to numerical diffusion) could possibly
explain this discrepancy \citep[cf.][]{Yeates:etal:2008}.}

By disregarding the observed sunspot latitudes in favor of an imposed
fixed latitude drift, the DGT model becomes affected by the
amplitude-dependent minimum shift. The extent to which this contributes
to the predictive skill of the model could be easily tested by replacing
the actual sunspot area data as input by a synthetic dataset with random
cycle amplitudes (without any memory), but keeping the correlation
resulting from the minimum shift of overlapping cycles.
\citet{Cameron:Schuessler:2007} used this approach to demonstrate the
effect on precursor methods. The amount of predictive skill shown by the
DGT model for such an input will directly indicate how strongly this
correlation affects the prediction.

\section{Discussion}

The shift of the solar minimum epoch due to the
amplitude-dependent asymmetry between the growth and decay phases of
solar cycles can explain the rather high correlation coefficients
between various precursor quantities and the strength of the subsequent
cycle \citep[e.g.,][]{Hathaway:etal:1999}. In this connection, we should
note that even a statistically highly significant correlation does not
automatically imply a high skill of the precursor for the actual {\em
prediction} of a future cycle:

\begin{enumerate}
\item A typical value of, say, $r=0.8$ for a `good' precursor leaves
  about one third of the variance in amplitude unexplained by the
  correlation, so that the prediction is prone to considerable
  statistical uncertainty. Even the value of $r=0.97$ shown by the method
  of \citet[][see Hathaway et al. 1999]{Thompson:1993} did not prevent
  an utterly inaccurate prediction for cycle 23 (30\% too high). 
\item The data used to determine the correlations in most cases cover
  only part of the known range of variability in the solar
  cycle. Periods such as the Maunder and Dalton minima indicate that
  nonlinear and stochastic effects may severely limit the predictability
  since it is unclear how strongly these effects are represented in the
  observed surface flux \citep{Bushby:Tobias:2007}. `Tuning' of free
  parameters in prediction schemes can lead to seemingly high
  correlations for past cycles while the predictive skill for future
  cycles in fact is much lower.
\end{enumerate}

The dependence of many precursor methods on knowing the minimum epoch
further limits their practical usefulness: the large relative
fluctuations of the activity indices around sunspot minimum make it
necessary to average the data (over 1 year, for instance) in order to
obtain a proper definition of the minimum epoch. Consequently, a
sensible prediction can only be made one or two years after
the minimum has passed, so that there is no big advantage compared 
to methods that use the information provided by the early phases of the
new cycle itself, like the steepness of the activity rise
\citep{Waldmeier:1936, Elling:Schwentek:1992}.

The relation between the shift of the minimum time and the amplitude
of the next cycle due to cycle overlapping is strongest for a dynamo where
the time interval between the onsets of activity of subsequent cycles is 
constant. Fluctuations in the timing of the onset of activity introduces 
noise into the relation between minimum shift and activity and thus reduces 
the predictive skill of precursor methods.  Quantifying such fluctuations
in the solar cycle data is difficult because the length of the available 
datasets is not sufficient to reliably determine the phase fluctuations and 
drifts \citep{Gough:1978,Hoyng:1996} and to decide on the existence or 
otherwise of a solar `clock' \citep[cf.][]{Dicke:1978, Dicke:1988}. 
However the expected shifts due to the overlapping of the cycles and 
the amplitude dependent asymmetry produce variations in the cycle length 
(as measured from minimum to minimum) which are consistent with those 
observed since the Dalton minimum. This means the effect we have outlined  
is relevant.  In passing, we also note that in the case of 
advection-dominated dynamo models the cycle period is determined by the 
meridional flow, {\bf which supports phase stability 
\citep{Charbonneau:Dikpati:2000, Charbonneau:2005}.}

For our understanding of the solar dynamo, it would certainly be useful
to clarify to which extent the skill of some prediction methods results
1) from correlations in the input data (such as discussed here), 2) from
capturing early high-latitude manifestations of an extended new cycle,
and 3) from capturing mechanisms that connect properties of the old
cycle to the strength of the new cycle (such as a dynamo model with a
memory of at least one cycle). It is reasonable to suppose that there
is some memory in the solar cycle i.e., that the solar cycle amplitudes do not
constitute a purely random sequence. Potentially important information
could be gleaned from identifying a quantity that unequivocally
represents this memory and is not `contaminated' by early information
leaking in from the new cycle. Clarifying whether the memory extends
over more than one cycle would help to decide between
diffusion-dominated and advection-dominated dynamo models
\citep{Yeates:etal:2008}.

\section{Conclusions}

We have confirmed a highly significant correlation between the growth
rate of activity during the early phase of a solar cycle and its maximum
amplitude for all global activity indices, i.e., Wolf and group sunspot
numbers, total sunspot area, and 10.7-cm radio flux. On the other hand,
there is no significant correlation between the decay rate in the late
cycle phase and the cycle amplitude.

Owing to the overlapping of individual cycles, this asymmetry leads to
an amplitude-dependent shift of the minimum epoch, thus explaining
(fully or partly) the predictive power of precursor methods which
(directly or indirectly) use the timing of the activity minimum as a
pivotal point. The resulting correlation in the sunspot area data
probably also affects the predictions with the dynamo-based model of
\citet{Dikpati:Gilman:2006}.

For our understanding of the origin of the solar magnetic field, it is
important to disentangle the effects of `real' physical precursors,
i.e., properties of the old cycle directly affecting the flux generation
for the next cycle or early high-latitude manifestations of the new
cycle, from apparent precursors, which derive their predictive power
from the the amplitude-dependent shift of the minimum epoch. 


\bibliographystyle{apj}

\begin{thebibliography}{46}

\expandafter\ifx\csname natexlab\endcsname\relax\def\natexlab#1{#1}\fi

\bibitem[{{Altrock} {et~al.}(2008){Altrock}, {Howe}, \&
  {Ulrich}}]{Altrock:etal:2008}
{Altrock}, R., {Howe}, R., \& {Ulrich}, R. 2008, in ASP Conf. Ser., Vol. 383,
  Subsurface and Atmospheric Influences on Solar Activity, ed. R.~{Howe}, R.~W.
  {Komm}, K.~S. {Balasubramaniam}, \& G.~J.~D. {Petrie}, 335

\bibitem[{{Altrock}(1997)}]{Altrock:1997}
{Altrock}, R.~C. 1997, \solphys, 170, 411

\bibitem[{{Balmaceda} {et~al.}(2005){Balmaceda}, {Solanki}, \&
  {Krivova}}]{Balmaceda:etal:2005}
{Balmaceda}, L., {Solanki}, S.~K., \& {Krivova}, N. 2005, Memorie della Societa
  Astronomica Italiana, 76, 929

\bibitem[{{Bushby} \& {Tobias}(2007)}]{Bushby:Tobias:2007}
{Bushby}, P.~J. \& {Tobias}, S.~M. 2007, \apj, 661, 1289

\bibitem[{{Cameron} \& {Sch{\"u}ssler}(2007)}]{Cameron:Schuessler:2007}
{Cameron}, R. \& {Sch{\"u}ssler}, M. 2007, \apj, 659, 801

\bibitem[{{Charbonneau}(2005)}]{Charbonneau:2005}
{Charbonneau}, P. 2005, Living Reviews in Solar Physics, LRSP-2005-2,
  http://solarphysics.livingreviews.org

\bibitem[{{Charbonneau} \& {Dikpati}(2000)}]{Charbonneau:Dikpati:2000}
{Charbonneau}, P. \& {Dikpati}, M. 2000, \apj, 543, 1027

\bibitem[{{Choudhuri} {et~al.}(2007){Choudhuri}, {Chatterjee}, \&
  {Jiang}}]{Choudhuri:etal:2007}
{Choudhuri}, A.~R., {Chatterjee}, P., \& {Jiang}, J. 2007, Phys. Rev. Lett.,
  98, 131103

\bibitem[{{Dicke}(1978)}]{Dicke:1978}
{Dicke}, R.~H. 1978, \nat, 276, 676

\bibitem[{{Dicke}(1988)}]{Dicke:1988}
---. 1988, \solphys, 115, 171

\bibitem[{{Dikpati} {et~al.}(2006){Dikpati}, {de Toma}, \&
  {Gilman}}]{Dikpati:etal:2006}
{Dikpati}, M., {de Toma}, G., \& {Gilman}, P.~A. 2006, Geophys. Res. Lett., 33,
  5102

\bibitem[{{Dikpati} {et~al.}(2008{\natexlab{a}}){Dikpati}, {de Toma}, \&
  {Gilman}}]{Dikpati:etal:2008b}
---. 2008{\natexlab{a}}, \apj, 675, 920

\bibitem[{{Dikpati} \& {Gilman}(2006)}]{Dikpati:Gilman:2006}
{Dikpati}, M. \& {Gilman}, P.~A. 2006, Astrophys. J, 649, 498

\bibitem[{{Dikpati} {et~al.}(2008{\natexlab{b}}){Dikpati}, {Gilman}, \& {de
  Toma}}]{Dikpati:etal:2008}
{Dikpati}, M., {Gilman}, P.~A., \& {de Toma}, G. 2008{\natexlab{b}}, \apjl,
  673, L99

\bibitem[{{Elling} \& {Schwentek}(1992)}]{Elling:Schwentek:1992}
{Elling}, W. \& {Schwentek}, H. 1992, \solphys, 137, 155

\bibitem[{{Gnevyshev}(1967)}]{Gnevyshev:1967}
{Gnevyshev}, M.~N. 1967, \solphys, 1, 107

\bibitem[{{Gough}(1978)}]{Gough:1978}
{Gough}, D. 1978, in Pleins Feux sur la Physique Solaire, ed. S.~{Dumont} \&
  J.~{Roesch} (CNRS, Paris), 81

\bibitem[{{Harvey}(1992)}]{Harvey:1992a}
{Harvey}, K.~L. 1992, in The Solar Cycle, ed. K.~L. Harvey (San Francisco:
  Astronomical Society of the Pacific, ASP Conf. Series Vol. 27), 335

\bibitem[{{Harvey}(1994)}]{Harvey:1994a}
{Harvey}, K.~L. 1994, in Solar Surface Magnetism, ed. R.~J. {Rutten} \& C.~J.
  {Schrijver} (Dordrecht: Kluwer), 347

\bibitem[{{Harvey} \& {White}(1999)}]{Harvey:White:1999b}
{Harvey}, K.~L. \& {White}, O.~R. 1999, \jgr, 104, 19759

\bibitem[{{Hathaway} {et~al.}(1994){Hathaway}, {Wilson}, \&
  {Reichmann}}]{Hathaway:etal:1994}
{Hathaway}, D.~H., {Wilson}, R.~M., \& {Reichmann}, E.~J. 1994, Sol. Phys.,
  151, 177

\bibitem[{{Hathaway} {et~al.}(1999){Hathaway}, {Wilson}, \&
  {Reichmann}}]{Hathaway:etal:1999}
---. 1999, J. Geophys. Res., 104, 22375

\bibitem[{{Hathaway} {et~al.}(2002){Hathaway}, {Wilson}, \&
  {Reichmann}}]{Hathaway:etal:2002}
---. 2002, Sol. Phys., 211, 357

\bibitem[{{Howe} {et~al.}(2006){Howe}, {Komm}, {Hill}, {Ulrich}, {Haber},
  {Hindman}, {Schou}, \& {Thompson}}]{Howe:etal:2006}
{Howe}, R., {Komm}, R., {Hill}, F., {Ulrich}, R., {Haber}, D.~A., {Hindman},
  B.~W., {Schou}, J., \& {Thompson}, M.~J. 2006, \solphys, 235, 1

\bibitem[{{Hoyng}(1996)}]{Hoyng:1996}
{Hoyng}, P. 1996, Sol. Phys., 169, 253

\bibitem[{{Hoyt} \& {Schatten}(1998)}]{Hoyt:Schatten:1998}
{Hoyt}, D.~V. \& {Schatten}, K.~H. 1998, Sol. Phys., 179, 189

\bibitem[{{Kane}(2007)}]{Kane:2007}
{Kane}, R.~P. 2007, \solphys, 243, 205

\bibitem[{{Kane}(2008)}]{Kane:2008}
---. 2008, \solphys, 248, 203

\bibitem[{{Lantos}(2000)}]{Lantos:2000}
{Lantos}, P. 2000, \solphys, 196, 221

\bibitem[{{Lantos}(2006)}]{Lantos:2006}
---. 2006, Sol. Phys., 236, 199

\bibitem[{{Lantos} \& {Richard}(1998)}]{Lantos:Richard:1998}
{Lantos}, P. \& {Richard}, O. 1998, Sol. Phys., 182, 231

\bibitem[{{Legrand} \& {Simon}(1981)}]{Legrand:Simon:1981}
{Legrand}, J.~P. \& {Simon}, P.~A. 1981, Sol. Phys., 70, 173

\bibitem[{{Li}(1999)}]{Li:1999}
{Li}, K. 1999, \aap, 345, 1006

\bibitem[{{Ohl}(1966)}]{Ohl:1966}
{Ohl}, A.~I. 1966, Soln. Dann., 12, 84

\bibitem[{{Podladchikova} {et~al.}(2008){Podladchikova}, {Lefebvre}, \& {van
  der Linden}}]{Podladchikova:etal:2008}
{Podladchikova}, T., {Lefebvre}, B., \& {van der Linden}, R. 2008, Journal of
  Atmospheric and Terrestrial Physics, 70, 277

\bibitem[{{Ramaswamy}(1977)}]{Ramaswamy:1977}
{Ramaswamy}, G. 1977, \nat, 265, 713

\bibitem[{{Schatten}(2003)}]{Schatten:2003}
{Schatten}, K.~H. 2003, Adv. Space Res., 32, 451

\bibitem[{{Schatten} {et~al.}(1978){Schatten}, {Scherrer}, {Svalgaard}, \&
  {Wilcox}}]{Schatten:etal:1978}
{Schatten}, K.~H., {Scherrer}, P.~H., {Svalgaard}, L., \& {Wilcox}, J.~M. 1978,
  Geophys. Res. Lett., 5, 411

\bibitem[{{Solanki} {et~al.}(2002){Solanki}, {Krivova}, {Sch{\" u}ssler}, \&
  {Fligge}}]{Solanki:etal:2002b}
{Solanki}, S.~K., {Krivova}, N.~A., {Sch{\" u}ssler}, M., \& {Fligge}, M. 2002,
  \aap, 396, 1029

\bibitem[{{Thompson}(1993)}]{Thompson:1993}
{Thompson}, R.~J. 1993, Sol. Phys., 148, 383

\bibitem[{{Waldmeier}(1935)}]{Waldmeier:1935}
{Waldmeier}, M. 1935, Mitt. Eidgen. Sternw. Z{\"u}rich, 14, 105

\bibitem[{{Waldmeier}(1936)}]{Waldmeier:1936}
---. 1936, Astron. Nachr., 259, 267

\bibitem[{{Waldmeier}(1955)}]{Waldmeier:1955}
---. 1955, {Ergebnisse und Probleme der Sonnenforschung.} (Leipzig, Geest {\&}
  Portig)

\bibitem[{{Wilson}(1994)}]{Wilson:1994}
{Wilson}, P.~R. 1994, {Solar and stellar activity cycles} (Cambridge
  Astrophysics Series, Cambridge University Press)

\bibitem[{{Wilson} {et~al.}(1988){Wilson}, {Altrock}, {Harvey}, {Martin}, \&
  {Snodgrass}}]{Wilson:etal:1988}
{Wilson}, P.~R., {Altrock}, R.~C., {Harvey}, K.~L., {Martin}, S.~F., \&
  {Snodgrass}, H.~B. 1988, \nat, 333, 748

\bibitem[{{Yeates} {et~al.}(2008){Yeates}, {Nandy}, \&
  {Mackay}}]{Yeates:etal:2008}
{Yeates}, A.~R., {Nandy}, D., \& {Mackay}, D.~H. 2008, \apj, 673, 544


\end{thebibliography}



\clearpage

\end{document}